\documentstyle[prl,aps,psfig]{revtex}
\tighten
\begin{document}
\draft
\twocolumn

\title{Critical Quantum Chaos in $2D$ Disordered Systems with Spin-Orbit
Coupling}

\author{G.N. Katomeris and S.N. Evangelou}
\address{ Department of Physics, University of Ioannina, 
Ioannina 45 110, Greece}

\date{\today}
\maketitle

\begin{abstract}

We examine the validity of the recently proposed 
semi-Poisson level spacing distribution function $P(S)$,
which characterizes `critical quantum chaos', 
in $2D$ disordered systems with spin-orbit coupling.
At the Anderson transition  we show that the semi-Poisson $P(S)$ 
can describe closely the critical distribution obtained with
averaged boundary conditions, over  
Dirichlet in one direction with periodic in the other and
Dirichlet in both directions.
We also obtain a sub-Poisson linear
number variance $\Sigma_{2}(E)\approx \chi_{0}+
\chi E$, with asymptotic value $\chi\approx0.07$. 
The obtained critical statistics,
intermediate between Wigner and Poisson,
is relevant for disordered systems and chaotic models.
\end{abstract}
\vspace{1cm}
In mesoscopic physics the effect of disorder on the electron
propagation leads to the zero-temperature quantum
Anderson metal-insulator transition,
which arises from the competition between quantum
tunelling and interference, as a function of disorder. 
For weak disorder the electrons diffuse, due to quantum tunelling,
and the system is metallic with correlated chaotic energy levels and 
`level-repulsion' described by Wigner statistics
\cite{sklov1,alts,guhr}. 
In the case of strong disorder the electrons localize
in random positions, due to quantum interference, 
and the system becomes insulating,
having non-chaotic completely uncorrelated (random) energy levels
which show `level attraction' and obey ordinary Poisson statistics. 
In order to see the metal-insulator transition
high enough space dimensionality 
(usually greater than $2$) is required and at 
the critical point, which corresponds to an
intermediate value of disorder, the level statistics changes
from Wigner to Poisson \cite{sklov1,alts}. The critical
electrons are neither extended nor localized and
it is believed that a new universal critical statistics, 
intermediate between Wigner and Poisson,  should apply
\cite {sklov2,kramer}. 
We aim to address the question of the critical statictics
in two dimensions ($2D$), where a metal-insulator transition
occurs in the presence of spin-orbit coupling \cite{evan}.

The stationary energy levels of
$2D$ quantum billiards (e.g. in the form of the stadium),
with zero potential inside and infinity outside,
can also display quantum chaotic behavior \cite{bohigas}.
The analogies in the level statistical description
bring together the two fields of mesoscopic physics and quantum chaos
and have been exploited in the past for understanding important
phenomena in both areas. In this respect, 
Wigner statistics  was originally  
conjectured to apply for quantum systems 
with chaotic classical dynamics, since the levels
resemble the eigenvalues
found in appropriate random matrix ensembles,
introduced long ago \cite{bohigas,mehta}. 
On the other hand, integrable systems correspond  
to Poisson statistics having completely uncorrelated (random)
eigenvalues. The key question is again what happens at criticality,
between chaos and integrability, similarly to
the transition between metal and insulator. 
Recently, a new distribution was proposed to describe
critical levels statistics \cite{bogomolny,braun}, 
which contains both Wigner and Poisson features, as the 
main theme of what is called `critical quantum chaos'. 
This intermediate distribution can be derived from a short range plasma
model \cite{bogomolny} and was realized in pseudo-integrable systems,
such as the classically non-integrable but of zero metric entropy
rational triangle billiards \cite{bogomolny,shudo}, 
and corresponds to other solvable models \cite{gaudin,yukawa}.
In disordered systems the intermediate distribution,
named semi-Poisson, was shown
to characterize critical states 
at the $3D$ metal-insulator transition \cite{braun} and
the energy levels of few electrons 
in the presence of interactions \cite{pichard}.

\par
\vspace{.4in}
\centerline{\psfig{figure=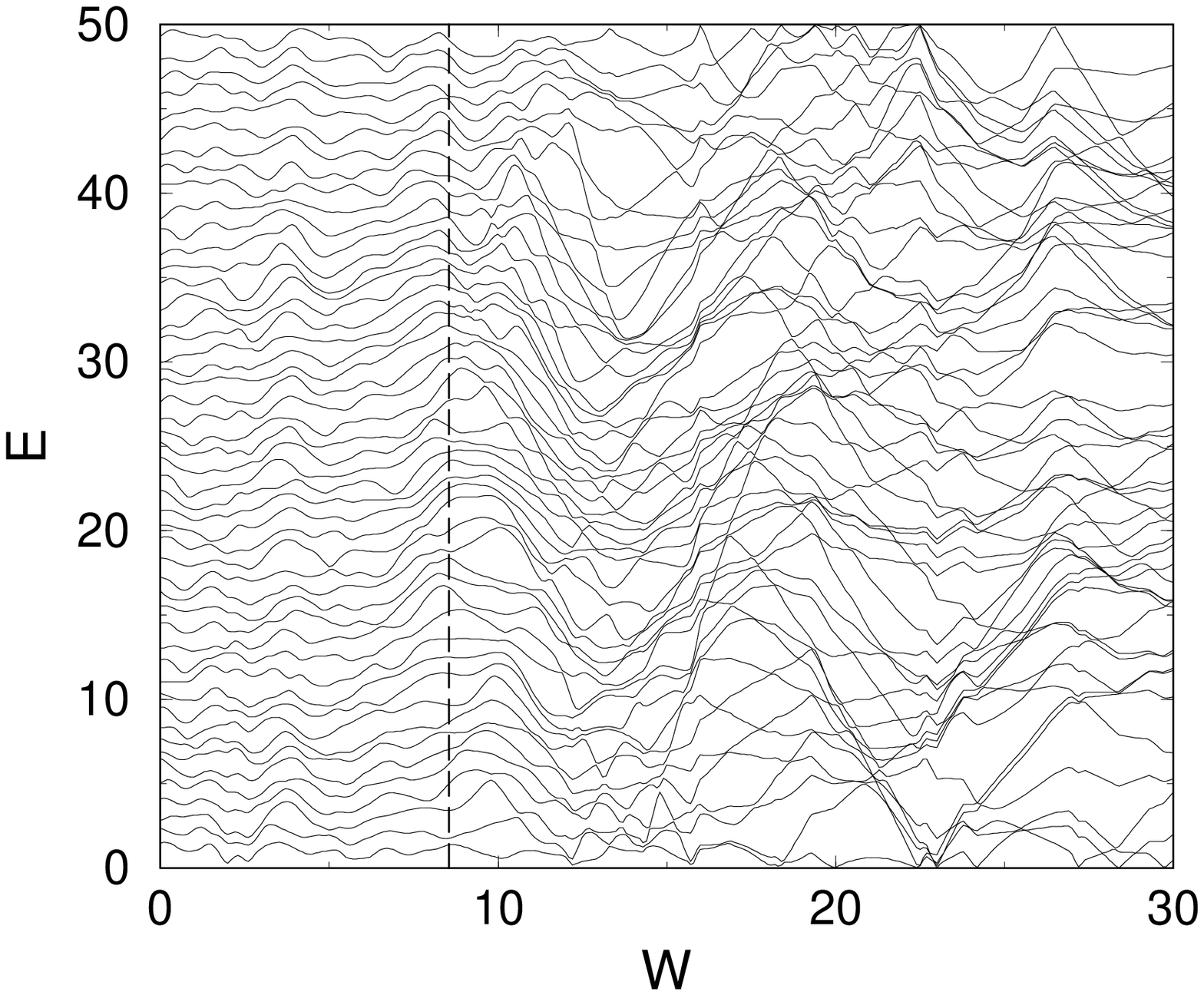,width=3.1in}}
{\footnotesize{{{\bf FIG. 1.}
Energy levels $E$ versus disorder $W$,
for a $2D$ disordered system with spin-orbit coupling.
The metal-insulator critical point $W_c=8.55$, marked with a
broken line, separates chaotic levels (on the left) with
`level-repulsion' and `spectral rigidity' described by Wigner statistics,
from non-chaotic levels (on the right) with `level attraction'
and `spectral randomness' described by Poisson statistics. 
In this paper we address what happens at the metal-insulator 
transition (broken line).
}
}}
\vspace{3mm}

In Fig.1 the energy levels $E$ obtained from our $2D$ disordered
system are displayed as a function of disorder $W$.  
The behavior of the levels is seen clearly
to change at the critical point of the metal-insulator transition $W_c$, 
which separates chaotic levels (on the left)
from non-chaotic levels (on the right). 
The chaotic levels are more regular (correlated) than
the non-chaotic levels,  which are uncorrelated (random).
The `level-repulsion' effect can be seen on the chaotic levels and
the `level-attraction' on the non-chaotic levels, where degeneracies 
exist. The displayed levels in Fig. 1
are obtained for one random configuration from
a system of linear size $L=30$, by keeping all the energies at the middle
of the spectrum, making the density of states constant at every $W$.
The overall `spectral rigidity', seen in the metallic chaotic levels,
can be contrasted
with the `spectral randomness' of the insulating non-chaotic levels.
The surprising result, also visualized in this figure,
is that the chaotic levels appear more `regular' than the non-chaotic ones.

The question addressed in this paper is: `what is the level
statistics at the critical broken line of Fig. 1?' 
This study is done in connection to the scenario of `critical 
quantum chaos', which is summarized in a semi-Poisson
level spacing distribution function $P(S)$ and a sub-Poisson linear 
number variance $\Sigma_{2}(E)$, which measures level-fluctuations 
in a given energy window $E$. 
Moreover, in order to examine the validity of the semi-Poisson $P(S)$ 
different boundary conditions (BC) must be considered, in the spirit
of recent important findings \cite{braun}. 
The fact that the critical level fluctuations are at the same time
scale-invariant (size-independent) and dependent on BC was explained
by invoking the concept of the critical conductance \cite{braun}.
The influence of boundary conditions on the critical level statistics,
without affecting the critical disorder, has also been demonstrated for a 
different critical $2D$ model \cite{schweitzer}.

The main features of `critical quantum chaos'
for systems characterized by the universality index $\beta=1,2,4$ are
summarized in:
(i) The semi-Poisson $P(S)$ level spacing distribution which  
shows Wigner-like repulsion $\sim S^{\beta}$ at small spacings $S<<1$
and is exponential, Poisson-like, 
$\sim \exp(-(\beta +1)S)$ at large spacings $S>>1$,
overall described by the scale-invariant normalized semi-Poisson curve
\begin{equation}
P(S)=A S^{\beta} \exp(-(\beta +1)S), 
\end{equation}
with the constant values $A=4$, $27/2$ and  $3125/24$ obtained
from normalization, respectively. The spacing distribution
$P(S)$, by applying a `level-unfolding' procedure 
which keeps the level-density constant, 
corresponds to  $\langle S \rangle =1$.
(ii) The sub-Poisson  number variance, which defines the level number
fluctuations in an energy window $E$, with the mean number
proportional to $E$ after `unfolding', 
according to this scenario is
\begin{equation}
\Sigma_{2}(E) \approx \chi_{0} + \chi E, 
\end{equation}
defining the level compressibility $\chi$. The value of $\chi$
ranges between $0$ (chaos) and $1$ (integrability) and
was related to the multifractality 
of the critical wavefunctions \cite{chalker}.

The considered disordered system displays a transition in $2D$ 
with energy levels which obey Wigner statistics for the metal  
(with $\beta=4$)
and  Poisson statistics for the insulator (see Fig. 1). 
At criticality, where one expects `critical quantum chaos' 
to apply,  numerical work  suggested  level-repulsion in $3D$
for small $S$ \cite{sklov2,kramer}, also later shown in $2D$ \cite{evan}.  
In order to study carefully the
level fluctuations in the critical region it is important to identify
the crucial role of BC \cite{braun}. We find that for the three
considered kinds of BC the critical
distribution function shows level repulsion at small spacings 
and is Poisson-like at large spacings.
However, when considering an averaged distribution over the cases:
1) Dirichlet BC in both directions and 2)
periodic BC in one direction and Dirichlet in the other, 
the obtained distribution is seen to be remarkably close to
the scale-invariant semi-Poisson curve of Eq. (1),
appropriate for $\beta=4$ (see Fig. 3 below).

The theoretical framework to study the Anderson transition can
classify disordered systems into
three universality classes, depending on whether the Hamiltonian preserves
the time-reversal invariance or the rotational invariance,
in direct analogy with the random matrix theory description 
of quantum chaotic systems \cite{bohigas,mehta}.
Zero spin-orbit corresponds to the orthogonal universality 
class ($\beta=1$) and finite spin-orbit breaks the rotational 
invariance so that one obtains the symplectic universality class ($\beta=4$).
In our calculations we consider a two-dimensional disordered system,
with spin-orbit coupling 
for  spin-${\frac {1} {2}}$ particles, described by the Hamiltonian
\cite{evan}
\begin{equation}
{\cal {H}} ={\Large \sum_{i,\sigma}} \epsilon_{i}
c_{i,\sigma}^{+} c_{i,\sigma} + {\Large \sum_{(i,j)}} 
{\Large \sum_{(\sigma,\sigma^{'})}}V_{i,j;\sigma,\sigma^{'}}
c_{i,\sigma}^{+} c_{j,\sigma^{'}},
\end{equation}
where $i$ labels the $L^{2}$ square lattice sites
and $\sigma=\pm 1/2 $ is the spin index on each site.
The  second sum is taken over all nearest neighbour lattice pairs $(i,j)$
and the random on site potential $\epsilon_{i}$ 
is a spin independent uniformly distributed random variable,
chosen from a probability distribution of width $W$. 
In this case the nearest neighbour hoppings $V_{i,j}$
are random $2\times 2$  matrices 
describing spin rotation, due to  spin-orbit,
on every lattice bond $(i,j)$. In the 
spinor space they  are represented by 
\begin{equation}
V_{i,j} =
\left(\begin{array}{ll}1+i\mu V^{z} & \mu V^{y}+i\mu V^{x}\\
-\mu V^{y}+i\mu V^{x} & 1-i\mu V^{z}
\end{array}\right)_{ij},
\end{equation}
where $\mu$ denotes the spin-orbit coupling and 
the $V^{x},V^{y}$ and $V^{z}$, defined for every bond $(i,j)$,
are real and independent random variables chosen
from a uniform probability distribution on the interval
$[-{\frac {1} {2}}, +{\frac {1} {2}}]$.  For the rest the
spin-orbit strength is fixed to $\mu=2$ and the disorder is chosen
to lie exactly at the critical point $W_{c}=8.55$ \cite{evan}.

We compute the eigenvalues from Eq. (3)
by diagonalizing numerically the corresponding Hamiltonian matrices
for large square lattices. 
The statistical analysis of energy levels must be done on
a constant density of states using an `unfolding procedure'.
In order to achieve the level unfolding for
the considered disordered system it is sufficient
to obtain the average of the integrated density of states
${\cal N}$, locally at $E$,
by repeating many times the disorder configuration
creating a statistical ensemble. 
Then the `raw' spacings $\Delta_{i}=E_{i}-E_{i-1}$ 
are replaced by the `unfolded' new ones $S_{i}={\cal N}_{av}(E_{i})-
{\cal N}_{av}(E_{i-1})\approx  (E_{i}-E_{i-1}) {\frac {\partial
{\cal N}_{av}(E)}{\partial E}}=(E_{i}-E_{i-1})/\Delta$,
where $\Delta$ is the local mean spacing around $E_{i}$
or equivalently the inverse density of states of the raw data.
In the numerical calculations
we considered eigenvalues within the energy window $[-2,2]$ performing
$2700, 1200, 675$ and $794$ random configuration
runs in each case, with $L=20, 30, 40$ and $60$, respectively.
The total number of eigenvalues from
all random configurations
for each BC is about $400,000$ for $L=20,30,40$ 
and $1,000,000$ for $L=60$.
These `raw' data were `unfolded' in the described way.
\par
\par
\vspace{.4in}
\centerline{\psfig{figure=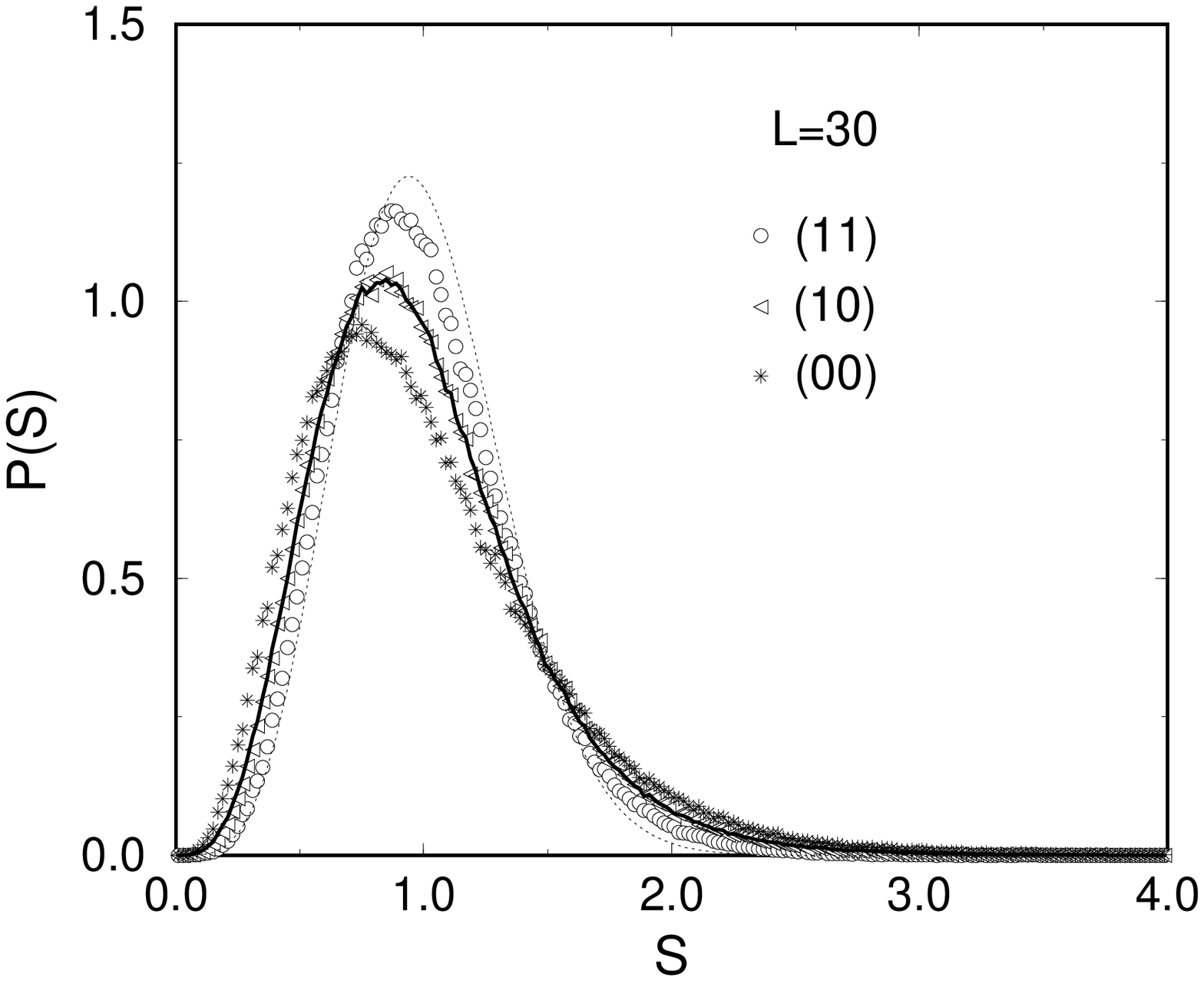,width=3.1in}}
{\footnotesize{{{\bf FIG. 2.}
This figure shows the variation of the critical $P(S)$
for three choices of  BC for a system of linear size $L$. 
The mean distribution over the three cases 
is shown by the continuous black line. The Wigner distribution (dotted
line) is also shown. 
}
}}
\vspace{3mm}

The obtained $P(S)$ at criticality is shown
in Fig. 2 for the three different choices of BC.
The computed curves are, clearly, very different,
in agreement with the corresponding $3D$ results 
for $\beta=1$ \cite{braun}.
They are very different from the Wigner or Poisson curves
while their average, over the three BC,
cannot fit to the semi-Poisson, either. However,
the average over Dirichlet (hard wall) in both directions $(00)$ and
periodic in one direction with Dirichlet in the 
other $(10)$, which is displayed in Fig. 3 for various system sizes, 
gives a distribution very well described by the semi-Poisson
curve of Eq. 1. This is the most important
result of the paper and shows the validity of
the semi-Poisson for the chosen specific average over BC
at criticality.  The obtained  semi-Poisson is also
in agreement with recent results for the critical
$P(S)$ at the metal-insulator transition in $3D$
disordered systems, where the semi-Poisson
was obtained by averaging over all possible combinations of BC \cite{braun}.
\par

\par
\vspace{.4in}
\centerline{\psfig{figure=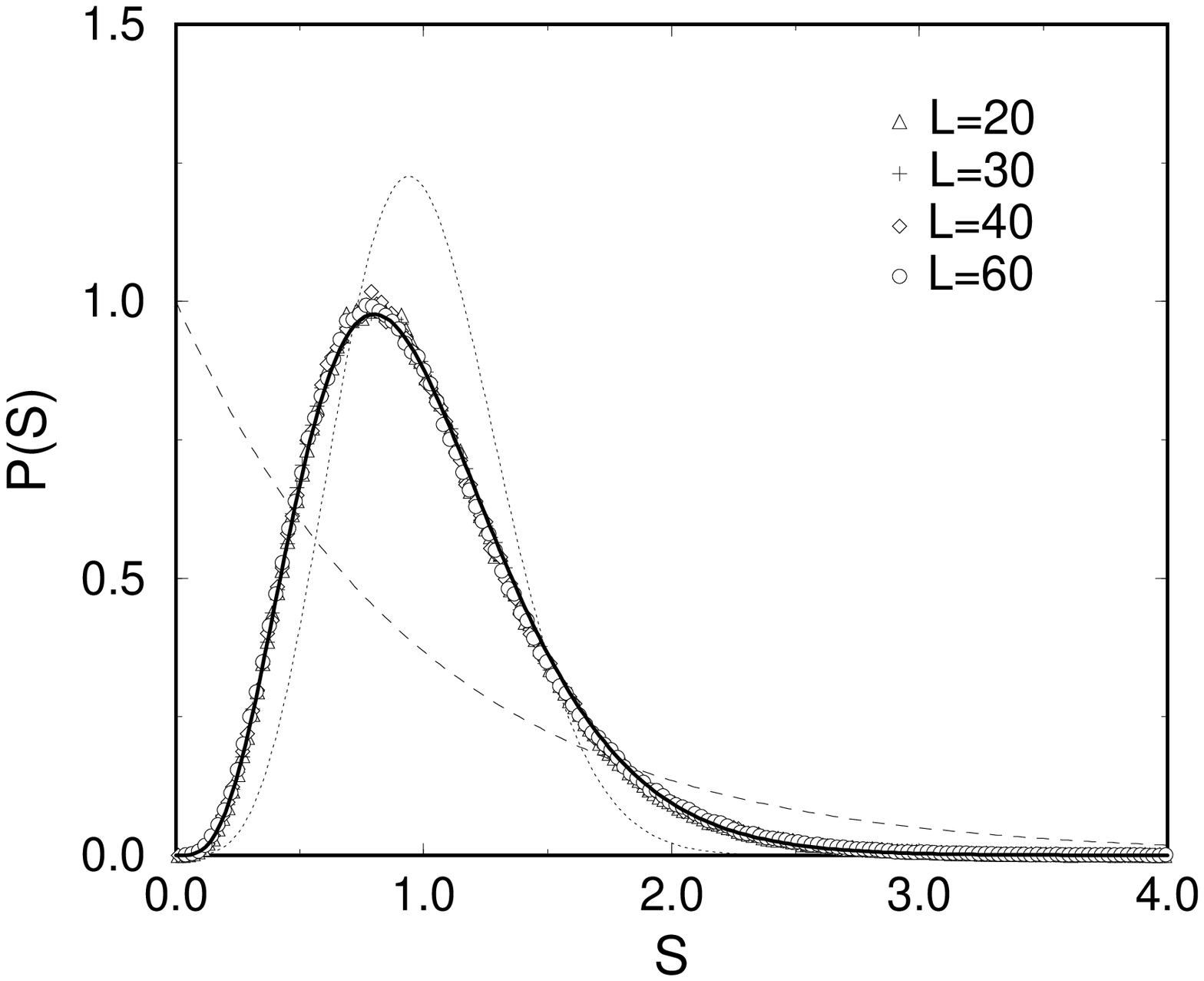,width=3.1in}}
{\footnotesize{{{\bf FIG. 3.}
The mean $P(S)$ distribution of the (10) and (00) combinations of BC
for several system sizes $L$
is shown to follow closely the semi-Poisson distribution 
$P(S)=(3125/24) S^{4} \exp(-5S)$ (Eq. (1) for $\beta=4$)
(black line).
The Wigner (dotted line) and the Poisson (dashed line) 
are also plotted.
}
}}
\par

\vspace{3mm}
The longer, in the $E$-range,
critical spectral fluctuations
are described by a linear number variance
$\Sigma_{2}(E)\approx \chi_{0}+\chi E$, with the compressibility $\chi$
related to the critical wavefunction dimension $D^{\psi}_2$
and the space dimension $d$ via $\chi = (1/2)(1-D^{\psi}_2 /d)$
\cite{chalker}.  
In the considered model previous
studies gave $D^{\psi}_2 \approx 1.63$ \cite{evan}.
For a rather small energy window $E$ (see Fig. 4) the
level number variance is shown to be linear
with level compressibility $\chi$ which varies with the chosen BC.
However, when the energy window $E$ increases the result becomes independent
of BC, as was already shown in $3D$ \cite{braun}. 
The obtained asymptotic level compressibility in this case
becomes $\Sigma_{2}(E)/E\to \chi \approx 0.07$ 
which leads to a $D_{2}^{\psi}$, rather close to the expected value
according to the previous formula.
The obtained value of $\chi$ in $2D$ should be contrasted with the higher $3D$
asymptotic value $\chi\approx 0.27$ \cite{braun}. 

The  main result from our calculations, done on a
square random lattice and not on a peculiarly shaped non-random
billiard, is the validity of the semi-Poisson statistics 
at the metal-insulator transition
in $2D$ disordered systems with spin-orbit coupling. 
However, the semi-Poisson $P(S)$
is obtained for the averaged distribution
over two specific BC. It must be pointed out that only
the main part of the distribution agrees surprisingly
well with the analytical result. For large $S$, where the dependence
on BC should become less important, we have not suceeded to
describe its tails, possibly due to their
exponentially small nature. In this case 
the appropriate statistical measure becomes the number
variance since longer range level correlations are needed.  
We find a linear number variance 
$\Sigma_{2} (E) \sim \chi E$ which becomes independent of the BC choice.
The obtained $\chi$ is close to the expected value from the
formula via $D^{\psi}_2$.
\par
\vspace{.4in}
\centerline{\psfig{figure=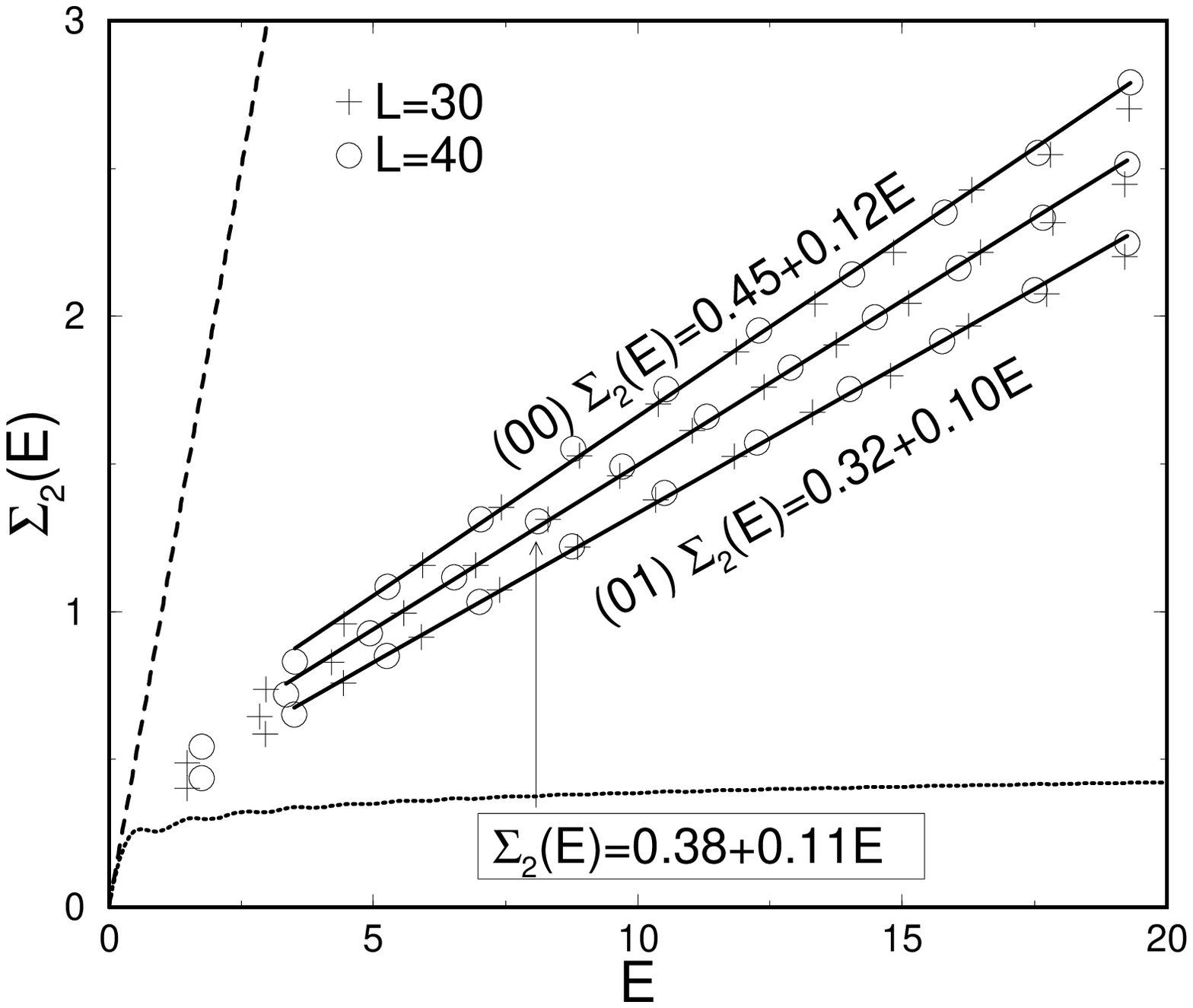,width=3.1in}}
{\footnotesize{{{\bf FIG. 4.}
The number variance $\Sigma_{2}(E)$ vs the energy window $E$ for the 
two BC (00) and (10). The straight lines fit the data giving slopes
corresponding to non-asymptotic $\chi$ values. 
The Wigner (dotted line) and the
Poisson (dashed line) are also displayed for comparison.
}
}}

\vspace{3mm}
In conclusion, we have shown the validity of the semi-Poisson 
level statistics at the critical point of the metal-insulator
transition with $\beta=4$ in $2D$. 
The semi-Poisson curve is shown to describe very well
the main part of the computed distribution for a specific
average over BC and
is similar to recent results for critical  disordered systems 
and weakly chaotic quantum systems.
Our calculations, on one hand, can give a justification 
for the averaging over boundary conditions
recently shown to lead to the semi-Poisson statistics
at the mobility edge \cite{braun}.
On the other hand, suggest that such an average
might be related to the bandwidth
distributions, by repeating periodically the square,
as it was recently shown for a non-random
one-dimensional critical quasi-periodic model \cite{pichard1}.
Clearly, more work is needed to examine the validity of the
critical semi-Poisson distribution, also for $3D$ disordered systems
with spin-orbit coupling and systems  
in the presence of a magnetic field ($\beta=2$).

This work was supported by a TMR network of EU
and a Greek-Chineese collaboration G.S.S.T.  
Useful discussions with J.-L. Pichard and G. Montambaux are 
gratefully aknowledged.
\par
\vspace{.4in}
\centerline{\psfig{figure=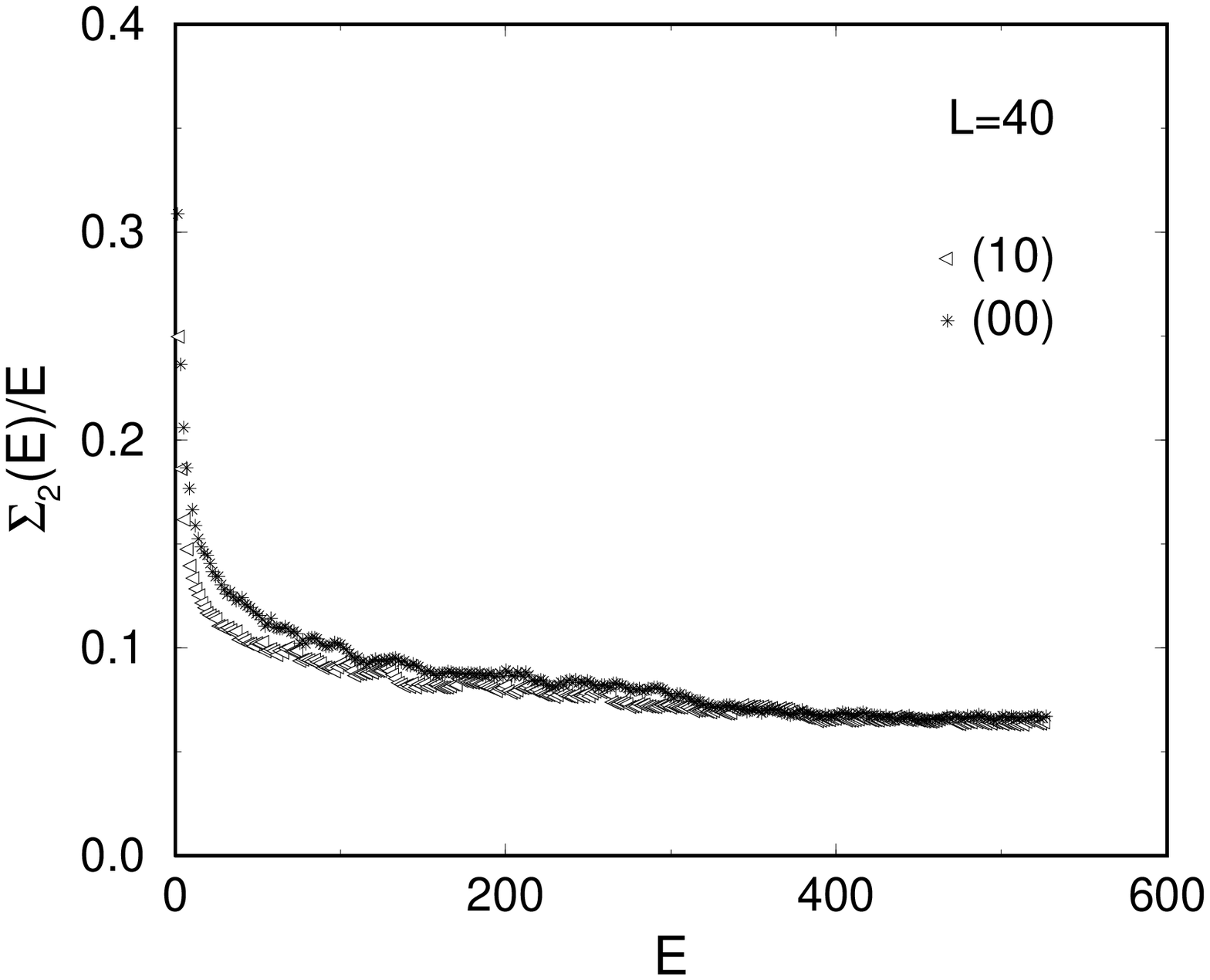,width=3.1in}}
{\footnotesize{{{\bf FIG. 5.}
$\Sigma_{2} (E)/E$ versus $E$ 
for a much broader range of $E$ where the independence on
BC is seen. The asymptotic value
approximates $\chi \approx 0.07$.
}
}}


\begin{references}

\bibitem{sklov1} Altshuler B.L. and Shklovskii B.I., 
Sov. Phys. JETP, {\bf 64}, 127 (1986).

\bibitem{alts} Altshuler B.L. {\it et} {\it al},
Sov. Phys. JETP {\bf 64}, 625 (1988).

\bibitem{guhr} Guhr T., M\"uller-Groeling A. and Weidenm\"uller H., 
Phys. Reports {\bf 229}, 189-425 (1998).

\bibitem{sklov2} Shklovskii B.I., Shapiro B., Sears B., 
Lambrianides P., and Shore H.B., Phys. Rev. B{\bf 47},  
11487 (1993).

\bibitem{kramer} Zharekeshev I. and Kramer B., 
Phys. Rev. Lett. {\bf 79}, 717  (1997) and references therein.

\bibitem{evan} Evangelou S.N. and Ziman T.A.L.,
J. Phys. C {\bf 20}, L235 (1987);
Evangelou S.N., Phys. Rev. Lett. {\bf 75}, 2550  (1995).

\bibitem{bohigas} Bohigas O. in Les Houches Summer School,
{\it Chaos and Quantum Physics}, eds. Giannoni M.J., 
Voros A., and  Zinn-Justin J., (Elsevier Science, 1991).

\bibitem{mehta} Mehta M.L.,
{\it Random Matrices}, (Academic Press, New York,
1991), 2nd ed.

\bibitem{bogomolny} Bogomolny E.B., Gerland U. and Schmit C.,
Phys. Rev. E {\bf 59}, R1315 (1999).

\bibitem{braun} Braun D., Montambaux G. and Pascaud M.,
Phys. Rev. Lett. {\bf 81}, 1062  (1998).

\bibitem{shudo} Shudo A. and Shimizu Y., Phys. Rev. E {\bf 47},
54 (1993).

\bibitem{gaudin} Gaudin M., Nuclear Phys. {\bf 85}, 545 (1966).

\bibitem{yukawa} Yukawa T., Phys. Rev. Lett. {\bf 54}, 1883 (1985).

\bibitem{pichard} Waintal X., Weinmann D. and Pichard J.-L.,
Eur. Phys. J. B{\bf 7}, 451 (1999).

\bibitem{schweitzer} Schweitzer L. and Potempa H.,
Physica A {\bf 266}, 486 (1999).

\bibitem{chalker}Chalker J.T., Kravtsov V.E., and Lerner I.V.,
JETP Lett. {64},  386 (1996); Kravtsov V.E. and Yudson V., 
Phys. Rev. Lett. {\bf 82}, 157 (1998).

\bibitem{pichard1} Evangelou S.N. and Pichard J.-L.,
cond-mat/ 9902321.

\end{references}
\end{document}